\documentclass[conference]{IEEEtran} % For conference Paper

\usepackage{hyperref} % create hyperlinks
\usepackage{graphicx}
\usepackage{amssymb}
\usepackage{amsmath}
\usepackage{mathtools}
\usepackage{cite}
\usepackage{stfloats}
\usepackage{subfigure}
\usepackage{psfrag}
\usepackage[mathscr]{euscript}
\usepackage{acronym}  % make an acronym
\usepackage{booktabs}
\usepackage[table]{xcolor}
\usepackage[utf8]{inputenc}
\usepackage[english]{babel}

\makeatletter
\renewcommand{\fnum@figure}{Fig. \thefigure}
\makeatother

\acrodef{IoT}{Internet of Thing}
\acrodef{D2D}{device-to-device}
\acrodef{BPP}{binomial point process}
\acrodef{MGF}{moment-generating function}
\acrodef{QoS}{quality of service}
\acrodef{QoP}{quality of performance}
\acrodef{GP}{geometric programming}
\acrodef{GGP}{generalized geometric programming}
\acrodef{ASE}{area spectral efficiency}
\acrodef{MISO}{multiple-input single-output}
\acrodef{KKT}{Karush Kuhn Tucker}
\acrodef{MAN}{multilayer aerial network}
\acrodef{AN}{aerial network}
\acrodef{UAV}{unmanned aerial vehicle}
\acrodef{STP}{successful transmission probability}
\acrodef{NLoS}{non-line of sight}
\acrodef{LoS}{line of sight}
\acrodef{MAN}{multi-layer aerial network}
\acrodef{U2U}{\ac{UAV}-to-\ac{UAV}}
\acrodef{U2G}{\ac{UAV}-to-ground}
\acrodef{SINR}{signal to interference plus noise ratio}
\acrodef{PGFL}{probability generating functional}
\acrodef{CCDF}{complementary cumulative distribution function}
\acrodef{CF}{characteristic function}
\acrodef{PPP}{Poisson point process}
\acrodef{CSI}{channel state information}
\acrodef{OFDM}{orthogonal frequency division multiplexing}
\acrodef{OFDMA}{orthogonal frequency division multiple access}

\acrodef{RV}{random variable}
\acrodef{i.i.d.}{independent, identically distributed}
\acrodef{PMF}{probability mass function}
\acrodef{PDF}{probability distribution function}
\acrodef{CDF}{cumulative distribution function}
\acrodef{ch.f.}{characteristic function}
\acrodef{AWGN}{additive white Gaussian noise}
\acrodef{SNR}{signal-to-noise ratio}
\acrodef{LRT}{likelihood ratio test}
\acrodef{DRT}{distance ratio test}
\acrodef{GLRT}{generalized likelihood ratio test}
\acrodef{CRLB}{Cram\'{e}r-Rao lower bound}
\acrodef{CRB}{Cram\'{e}r-Rao bound}
\acrodef{ZZLB}{Ziv-Zakai lower bound}
\acrodef{ZZB}{Ziv-Zakai bound}
\acrodef{LOS}{line-of-sight}
\acrodef{ToF}{time-of-flight}
\acrodef{NLOS}{non-line-of-sight}
\acrodef{GDOP}{geometric dilution of precision}
\acrodef{GPS}{Global Positioning System}
\acrodef{FIM}{Fisher information matrix}
\acrodef{PEB}{position error bound}
\acrodef{SPEB}{squared position error bound}
\acrodef{TOA}{time-of-arrival}
\acrodef{TOF}{time-of-flight}
\acrodef{WSN}{wireless sensor network}
\acrodef{MAC}{medium access control}
\acrodef{RSS}{received signal strength}
\acrodef{WAF}{wall attenuation factor}
\acrodef{TDOA}{time difference-of-arrival}
\acrodef{RF}{radiofrequency}
\acrodef{RTT}{round-trip time}
\acrodef{AOA}{angle-of-arrival}
\acrodef{MF}{matched filter}
\acrodef{ED}{energy detector}
\acrodef{ML}{maximum likelihood}
\acrodef{MSE}{mean-square error}
\acrodef{RMSE}{root-mean-square error}
\acrodef{LEO}{localization error outage}
\acrodef{ppm}{part-per-million}
\acrodef{ACK}{acknowledge}
\acrodef{UWB}{Ultrawide bandwidth}
\acrodef{TNR}{threshold-to-noise ratio}
\acrodef{LS}{least squares}
\acrodef{IR-UWB}{impulse radio UWB}
\acrodef{FCC}{Federal Communications Commission}
\acrodef{TH}{time-hopping}
\acrodef{PPM}{pulse position modulation}
\acrodef{MUI}{multi-user interference}
\acrodef{PDP}{power delay profile}
\acrodef{BPZF}{band-pass zonal filter}
\acrodef{SIR}{signal-to-interference ratio}
\acrodef{RFID}{radio frequency identification}
\acrodef{WPAN}{wireless personal area network}
\acrodef{WWB}{Weiss-Weinstein bound}
\acrodef{DP}{direct path}
\acrodef{MF}{matched filter}
\acrodef{MMSE}{minimum-mean-square-error}
\acrodef{SBS}{serial backward search}
\acrodef{SBSMC}{serial backward search for multiple clusters}
\acrodef{NBI}{narrowband interference}
\acrodef{WBI}{wideband interference}
\acrodef{INR}{interference-to-noise ratio}
\acrodef{CR}{channel response}
\acrodef{CIR}{channel impulse response}
\acrodef{CR}{channel  response}
\acrodef{RADAR}{radar}
\acrodef{MUR}{Multistatic radar}
\acrodef{JBSF}{jump back and search forward}
\acrodef{HDSA}{high-definition situation-aware}
\acrodef{RRC}{root raised cosine}
\acrodef{ST}{simple thresholding}
\acrodef{BTB}{Bellini-Tartara bound}
\acrodef{P-Max}{$P$-Max}  %suggestion, use with \acl{P-Max}
\acrodef{MIMO}{multiple-input multiple-output}
\acrodef{MAP}{maximum a posteriori}
\acrodef{FG}{factor graph}
\acrodef{OP}{outage probability}
\acrodef{WED}{wall extra delay}
\acrodef{RMS}{root mean square}
\acrodef{SPAWN}{sum-product algorithm over a wireless network}
\acrodef{MDD}{minimum distance distribution}
\acrodef{MAP}{maximum a posteriori probability}
\acrodef{PAR}{probabilistic association rule}

\usepackage{color}
\usepackage{dsfont}
\usepackage{bbm}

\def\cb{c_\text{o}}
\newcommand{\normx}{\left\lVert \bold{x}-\bold{x}_\text{rec} \right\rVert}

\newcommand{\jcy}{\bx\ijCo(y)}
\newcommand{\bx}{\mathbf{x}}
\newcommand{\ijCo}{_{j}^{(c)}}

\newcommand{\ikTo}{_{k}^{(\cb)}}
\newcommand{\ijcoL}{_{j}^{\text{(L)}}}
\newcommand{\ijcoN}{_{j}^{\text{(N)}}}
\newcommand{\Co}{^{(c)}}
\newcommand{\To}{^{(\cb)}}
\newcommand{\CoTo}{^{(c, \cb)}}
\newcommand{\RI}{R\ijk}
\newcommand{\ijk}{_{j,k}\CoTo}
\newcommand{\ijkC}{_{j,k}\Co}
\newcommand{\coL}{^{\text{(L)}}}
\newcommand{\coN}{^{\text{(N)}}}

\newcommand{\hij}{h_{j}}
\newcommand{\hik}{h_{k}}

\newcommand{\aCo}{^{-\alpha\Co}}
\newcommand{\aTo}{^{-\alpha\To}}
\newcommand{\acoN}{^{-\alpha\coN}}
\newcommand{\acoL}{^{-\alpha\coL}}

\newcommand{\Lap}[1]{\mathcal{L}_{#1}}

\newcommand{\pijc}{p_{j}^{(c)}}

\newcommand{\Tin}{\cb\in\text{\{L,N\}}}

\newcommand{\Ws}[2]{{W_{}^{}}} % Symbol bandwidth
\newcommand{\TSIR}[2]{{\tau_{}^{}}}

\DeclareMathAlphabet{\mathsf}{OML}{cmbr}{m}{it}

\newtheorem{lemma}{Lemma}
\newtheorem{corollary}{Corollary}

\newcommand{\bd}{\begin{description}}
\newcommand{\ed}{\end{description}}
\newcommand{\be}{\begin{enumerate}}
\newcommand{\ee}{\end{enumerate}}
\newcommand{\bi}{\begin{itemize}}
\newcommand{\ei}{\end{itemize}}
\newcommand{\bl}{\begin{list}}
\newcommand{\el}{\end{list}}
\newcommand{\bt}{\begin{tabbing}}
\newcommand{\et}{\end{tabbing}}

\newcommand{\red}[1]{{#1}}

\newcommand{\blue}[1]{{#1}}

\acrodef{BS}{base station}
\acrodef{AP}{access point}
\begin{document}

\newcommand{\paperTitle}{
{Performance Analysis for\\ Multi-layer Unmanned Aerial Vehicle Networks}
%Multi-layer Aerial Networks:\\Modeling and Performance Analysis
}

% paper title
\title{\paperTitle}
% \title{Distributed Secrecy in \\Multilevel Wireless Networks}

% author names, IEEE memberships, corresponding address, title footnote %
\author{
\IEEEauthorblockN{
        Dongsun Kim\IEEEauthorrefmark{2},
        Jemin~Lee\IEEEauthorrefmark{2}, 
        and 
        Tony Q. S. Quek\IEEEauthorrefmark{1} 
}\\[-0.5em]
\IEEEauthorblockA{
\IEEEauthorrefmark{2}%Department of Information and Communication Engineering (ICE)\\
Daegu Gyeongbuk Institute of Science and Technology (DGIST), Korea\\
\IEEEauthorrefmark{1}Singapore University of Technology and Design (SUTD), Singapore\\
Email: {yidaever@dgist.ac.kr}, {jmnlee@dgist.ac.kr}, tonyquek@sutd.edu.sg
}
}
\maketitle %% make the title area
%\newpage
\setcounter{page}{1}
\acresetall
%%---------------------------------------------------------------------------%
%%                           abstract and key words                          %
%%---------------------------------------------------------------------------%
\begin{abstract}
%The \ac{UAV} has been introduced for the \ac{BS} which have difference in mobility and channel compared with the terrestrial \ac{BS}.
% In the wireless network, the use of the \ac{UAV} can take advantage of flexibility and mobility.
% and \ac{LoS} channels.
%In this paper, we provide the model of the \ac{MAN}, composed \acp{UAV} that distributed in \acp{PPP} with different densities, floating heights, and transmission power. In our model, we consider the \ac{LoS} and \ac{NLoS} channels and the probability of forming \ac{LoS}.
In this paper, we provide the model of the \ac{MAN}, composed \acp{UAV} that distributed in \ac{PPP} with different densities, heights, and transmission power. In our model, we consider the \ac{LoS} and \ac{NLoS} channels \red{which is probabilistically formed.}
We firstly derive the \ac{PDF} of the main link distance and the Laplace transform of interference of \ac{MAN} considering strongest average received power-based association.
%
%We then analyze the \ac{STP} of the \ac{MAN}.
%We also provide the upper bound of the optimal density that maximizes the \ac{STP}.
\red{We then analyze the \ac{STP} of the \ac{MAN} and provide the upper bound of the optimal density that maximizes the \ac{STP} of the \ac{MAN}.}
%after deriving the \ac{PDF} of the main link distance and the Laplace transform of interference considering strongest transmitter selection. Addition, we derive the upper bound of the optimal density that maximizes the \ac{STP} by analyzing the range of the density of the \ac{AN}.
Through the numerical results, we show the existence of the optimal height of \ac{UAV} due to a performance tradeoff caused by the height of the \ac{AN}, and also show the upper bounds of the optimal densities in terms of the \ac{STP}, which decrease with the height of the \acp{AN}.
%For the case of \ac{MAN} with fixed total density of \acp{UAV}, we also discuss how the \acp{UAV} need to be distributed in different layers.

%In addition, we give the result of the 2-layer \ac{MAN} that shows the optimal densities which shows that when the total density of the \ac{UAV} is large, better to increase the number of the lower \ac{UAV}, and vise versa.

%In the wireless network, the use of \ac{UAV} takes advantage of flexibility and higher performance due to the mobility and \ac{LoS} channels.
%In this paper, motivated by the recent advances in \ac{UAV} wireless networks, we provide analysis for \ac{MAN}, composed of multi-layer networks with a \acp{UAV} distributed in \ac{PPP} with different densities, transmission powers, and altitudes for the communications from \acp{UAV} to theground receivers.
%Especially, we analyze the \ac{STP} for both when the main link distance is given and when \acp{UAV} communicate to the nearest receivers.
%% which gives \ac{ASE} of networks.
%By quantifying the \ac{STP} together with \ac{ASE},
%we explore the effects of altitudes and densities of \acp{AN} and provide that the \ac{STP} can be enhanced by controlling altitudes of \acp{AN}, which are useful for designing wireless networks involving \acp{UAV}.

\begin{IEEEkeywords}
Stochastic geometry, aerial networks, multi-layer, Poisson point process, association rule, optimal density.
\end{IEEEkeywords}
\end{abstract}

%\clearpage
\acresetall
%%%%%%%%%%%%%%%%%%%%%%%%%%%%%%%%%%%%

%%---------------------------------------------------------------------------%
%%                           Sec: Introduction                                %
%%---------------------------------------------------------------------------%

\section{Introduction}

Recent developments in the \ac{UAV} technology increase payloads capacity, average flight time, and battery capacity that enables the \ac{UAV} to play an important role in wireless networks.
In the area which needs quick deployment of the \ac{BS} due to disaster or events,
\acp{UAV} are expected to act as a temporal \ac{BS} as well \cite{ZenZhaLim:16}.
Furthermore, the data collection from the devices under certain energy constraints can be done by using \acp{UAV} \cite{HuaNisKat:14}.
In addition, demands on the data acquisition using \ac{UAV} in crowd surveillance have arisen \cite{MotBagTal:17}.
To utilize \acp{UAV} for the aforementioned applications and services, the research on the establishment of reliable \ac{AN} is required.

%The challenge for the aforementioned works \cite{ZenZhaLim:16,HuaNisKat:14,MotBagTal:17} are related to the wireless networks consists of \acp{UAV}.
%Hence, research on the wireless communication and the network composed of \ac{UAV} is necessary to enable and utilize \ac{UAV} aided services.

%Wireless networks involving the \ac{UAV} has been studied considering the \ac{LoS} channel probability \cite{AlKanJam:14, AlKanLar:14}, the pathloss and the fading of U2G channel \cite{AlKanJam:14}, the \ac{UAV} use as relay \cite{ GuaDevWan:14,MozSaaBen:16}.
%In \cite{AlKanJam:14}, the authors model the pathloss and the channel gain between   \ac{UAV} and ground channel by ray tracing in an urban area.
%Similarly, in \cite{AlKanLar:14}, they modeled the \ac{LoS} probability, which is determined by the angle from the ground, and proposed the optimal \ac{UAV} deployment that maximizes the coverage area.
%The research on \ac{UAV} use as relay networks in device-to-device communication and cellular networks is in \cite{ MozSaaBen:16 } and \cite{ GuaDevWan:14}, respectively.
%However, \cite{AlKanLar:14, MozSaaBen:16} did not consider the interference from the other \acp{UAV} and \cite{GuaDevWan:14} need more general scenario which considered the interference.

The \ac{UAV} based wireless communications
has been studied in \cite{AlKanLar:14,GuoDevWan:14,MozSaaBen:16} after modeling
the wireless channel and the mobility, which are different from those of the terrestrial networks.
%, which gives the opportunity to have better performance.
%The LoS channel is formed by the less obstruction that gives a smaller power attenuation and more deterministic channel gain.
%In \cite{AlKanJam:14}, the pathloss and the channel gain of the link between \ac{UAV} and ground node are modeled by regression of ray tracing in urban area.
In \cite{AlKanLar:14}, the probability that a link forms \ac{LoS}, i.e., the \ac{LoS} probability, is modeled, which is determined by the angle from the ground, and also proposed the optimal \ac{UAV} deployment that maximizes the coverage area.
In addition, \ac{UAV} relay networks in cellular networks and device-to-device communications are considered in \cite{GuoDevWan:14} and \cite{MozSaaBen:16}, respectively.
However, the studies mentioned above % \cite{AlKanJam:14,AlKanLar:14,GuaDevWan:14,MozSaaBen:16}
have considered only the small number of \acp{UAV}, which can show the performance only for the limited scenarios. %of \acp{UAV} which is unrealistic due to the spectrum sharing between \ac{UAV}.
%Since \acp{UAV} can share frequency resource, the interference should be considered.

%Recently, the research on the \acp{AN}, which consist of \acp{UAV}, is presented in \cite{ZhaZha:16,CheDhi:17,AzaRosChi:17}
%using
%stochastic geometry, which is a widely-used tool for randomly distributed nodes\cite{HaeGan:08}.
%The coexistence of \ac{AN} and the terrestrial cellular networks is studied by considering
%the distribution of \acp{UAV} as 3-D \ac{PPP} and 2-D \ac{PPP} in \cite{ZhaZha:16} and \cite{AzaRosChi:17}, respectively.
%The coverage probability of \ac{UAV} by using \ac{BPP}-based distribution is presented in \cite{CheDhi:17}.
%In addition, \cite{SekTabHos:18} studies the multi-tier drone network and shows the downlink spectral efficiency of the network via simulations.
%However, except for \cite{SekTabHos:18}, most of these works did not consider the multiple layer structure of \acp{AN}, of each layer has \acp{UAV} in different heights.
%In \acp{AN}, we can have various types of \acp{UAV} with different floating heights and transmission power depending on their hardware constraints\cite{ChaGomAl:16}. In addition, as the number of \acp{UAV} increases, it is required to have the \emph{multiple layer structure} in \ac{AN} for efficient resource management and reliable coexistence among \acp{UAV}.
%In \cite{SekTabHos:18}, the multiple layer of \acp{AN} is considered, but no performance analysis has been provided.
Recently, the research on the \acp{AN}, which consist of \acp{UAV}, is presented in \cite{CheDhi:17,AzaRosChi:17}
using
stochastic geometry, which is a widely-used tool for randomly distributed nodes\cite{HaeGan:08}.
The coexistence of \ac{AN} and the terrestrial cellular networks is studied by considering
the distribution of \acp{UAV} as \ac{PPP} in \cite{AzaRosChi:17}.
The coverage probability of \ac{UAV} by using \ac{BPP}-based distribution is presented in \cite{CheDhi:17}.
In addition, \cite{SekTabHos:18} studies the multi-tier \ac{UAV} networks and shows the downlink spectral efficiency of the network via simulations.
\red{However, except for \cite{SekTabHos:18}, most of these works did not consider the multiple layer structure of \acp{AN}. Nevertheless, no analytical approach is provided in \cite{SekTabHos:18}. }%, of each layer has \acp{UAV} in different heights.
%In \acp{AN}, we can have various types of \acp{UAV} with different heights and transmission power depending on their hardware constraints \cite{ChaGomAl:16}. In addition, as the number of \acp{UAV} increases, it is required to have the \emph{multiple layer structure} in \ac{AN} for efficient resource management and reliable coexistence among \acp{UAV}.
\blue{
In \acp{AN}, the \ac{UAV} have limitation on height due to the hardware and the law \cite{ChaGomAl:16}.
Furthermore, as a number of \ac{UAV} have mobility for serving different service, to avoid the collision between \acp{UAV} and to efficiently manage the resource, it is required to have the \emph{multiple layer structure} in \ac{AN} which differentiate the height according to the roles and types of \ac{UAV}.
}
%In \cite{SekTabHos:18}, the multiple layer of \acp{AN} is considered, but no performance analysis has been provided.

Therefore, in this paper, we investigate the performance of the \ac{MAN} with various types of \ac{UAV}.
% in different heights.
Specifically, the \ac{MAN} is composed of $K$ layer \acp{AN} that have \acp{UAV} with different transmission power, spatial densities, and heights.
Note that the multiple layer structure has been considered for terrestrial networks, which is called as the heterogeneous networks \cite{SinDhiAnd:13,DhiGanBacAnd:12, ZhaYanQueLee:17}.
However, different to those works, the heights of nodes need to be considered together with the channel model which has the \ac{LoS} probability determined by the height of node. This leads to new analysis on the interference and the \ac{STP} of the \ac{MAN} considering association rule.
\blue{To our best knowledge, there is no analysis on the \ac{STP} for multi-layer \ac{UAV} networks considering association rule, \ac{LoS}, and \ac{NLoS} channel. Furthermore, our analysis on the upper bound of the optimal density gives useful insights for future \ac{MAN} implementation.}
%
%Therefore, we consider the \ac{MAN} that composed of the \ac{AN} that have different power, density and height which is synonym of the heterogeneous wireless network in terrestrial networks \cite{SinDhiAnd:13,DhiGanBacAnd:12, ZhaYanQueLee:17}.
%Both works consider the communication considering the association for the \ac{BS} which have different power and densities.
%Compared with the \cite{SinDhiAnd:13,DhiGanBacAnd:12}, however, consideration on the height is inevitable in the \ac{AN} which demands new approach of analysis in aspect to the association and \ac{STP} analysis due to the \ac{LoS} and \ac{NLoS} channel of the \ac{UAV} communication.
%
Our contribution can be summarized as follows:
\begin{itemize}
\item using stochastic geometry, we newly analyze the Laplace transform of interference of \ac{MAN} by considering
\ac{NLoS} and \ac{LoS} channels with the height-dependent \ac{LoS} probability;
\item we derive the \ac{STP} when a ground node selects a \ac{UAV} with association rules considering strongest average received power for communications;
\item to give insight on the effect of \ac{UAV} density on \ac{STP}, we provide the upper bounds of the optimal \ac{UAV} densities of each layer \ac{AN} that maximize the \ac{STP}; and
\item we show the effects of channel and network parameters on the optimal heights of \acp{AN} and the compatibility of the upper bound of optimal densities.
\end{itemize}

\section{System Model}
%%---------------------------------------------------------------------------%
%%                           Sec: Network Model                               %
%%---------------------------------------------------------------------------%
In this section, we present the system model of \ac{MAN} with \ac{UAV} including the network description and the channel model. Furthermore, we describe the association rule which is used to obtain the \ac{PDF} of the main link distance.

\subsection{Multi-layer Aerial Networks}

\begin{figure}[t!]
    \begin{center}
    {
	 \includegraphics[width=1.00\columnwidth]{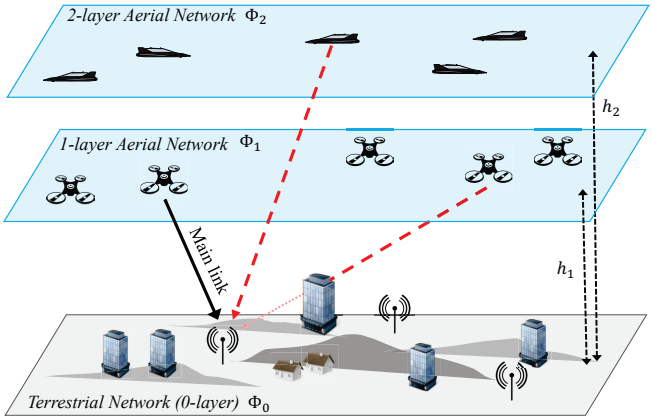}
   \vspace{-8mm}
    }
    \end{center}
    \caption{
    An example of two layer \ac{AN} with ground nodes (i.e., $0$-layer).
   % Transmission in the \ac{MAN} consists of dual \acp{AN} and ground receivers.
    The black lines represent the main link from a transmitter to a receiver and red dashed lines represent interference links which comes from other \acp{UAV}.
     % \vspace{-6mm}
   % The sets of the locations of \acp{UAV} and ground nodes are, respectively, $\Phi_0$, $\Phi_1$, and $\Phi_2$.
		}   \label{fig:System}
\end{figure}
% in different heights which is displayed in Fig~.\ref{fig:System}.
% MAN general

We consider a \ac{MAN} which consists of $K$ layers of \acp{AN} at different altitudes
with a terrestrial network
as shown in Fig.~\ref{fig:System}.
We denote $\mathcal{K}$ as the set of \ac{AN} layer indexes, i.e., $\mathcal{K}=\{1, \cdots, K\}$,
and layer $0$ as the terrestrial network.
We assume \ac{UAV} in \acp{AN} and the ground nodes in the terrestrial network are distributed according to \acp{PPP} \cite{HaeGan:08}. 
\blue{Although each \ac{UAV} has given path scheduled by controller, location of \acp{UAV} at given time is random from the perspective of other layers, hence, we use the \ac{PPP} for the location of the \ac{UAV} as \cite{AzaRosChi:17}.}
Specifically, in the $k$-layer, the node locations follows a homogeneous \ac{PPP} $\Phi_k$ with density $\lambda_k$ and they are at the fixed altitude $h_k$ and transmit with the power $P_k$.
Note that the altitude of nodes in the $0$-layer (i.e., the terrestrial layer) is $h_0 = 0$ and altitudes of other layers are $h_k >0$ for $k \in \mathcal{K}$.

%%We assume locations of the \ac{UAV} and the ground \ac{BS} follow a homogeneous \ac{PPP} $\Phi_k$ with density $\lambda_k$, and they are flying at the fixed altitude $h_k$ and transmit with power $P_k$.
%%Though, we assume that power of the \ac{UAV} is same, e.g., $P_k=P$.
%%We denote ground \acp{BS} as $0$-layer and use as the receiver.
%%However, expansion, e.g., use of ground \ac{BS} as a transmitter, is possible.
%%
%%
%
%We denote $\mathcal{K}$ as the set of layers, i.e., $\mathcal{K}=\{1, \cdots, K\}$,
%and its pair of symbols $ij$ ($i$, $j$ $\in\mathcal{K}$) to present the value or function when a transmitter is in the $i$-layer and a receiver is at the $j$-layer.
%%
%We assume locations of the \ac{UAV} and the ground \ac{BS} follow a homogeneous \ac{PPP} $\Phi_k$ with density $\lambda_k$, and they are flying at the fixed altitude $h_k$ and transmit with power $P_k$.
%Though, we assume that power of the \ac{UAV} is same, e.g., $P_k=P$.
%We denote ground \acp{BS} as $0$-layer and use as the receiver.
%However, expansion, e.g., use of ground \ac{BS} as a transmitter, is possible.

In the \ac{MAN}, we consider the communication from a \ac{UAV} to a ground node.\footnote{\blue{Note that we omitted analysis including the air-to-air channel in the paper, even though our framework is readily expandable for communication in between \ac{UAV} which will be presented in the future research.}}
%In the \ac{MAN}, we focus on the communication between ground \ac{BS} and \ac{UAV} but easily extended to the communication between \acp{BS} and between \acp{UAV}.
%In the traditional communication, as obstructions between ground \ac{BS} frequent which results in \ac{NLoS} channel that gives severe power attenuation and channel randomness.
%
In the communication with \acp{UAV},
%In the communication communication between \ac{UAV} and ground \ac{BS},
we should consider both \ac{LoS} and \ac{NLoS} channels since the existence of obstacles (e.g., buildings) between the transmitter and the receiver can be changed with the altitude of \ac{UAV}.
In \cite{AlKanLar:14}, the probability of forming \ac{LoS} chanel is modeled
by a signomial approximation of the probability of having obstructions between transmitter and receiver.
When a node in the $k$-layer transmits to a ground node, the \ac{LoS} probability is defined as\cite{AlKanLar:14}
\begin{align}
	\rho_{k}^{\text{(L)}}\left(x\right)=
	\frac{1}{1+a\exp(-b[\sin^{-1}\left(\frac{\hik}{x}\right)-a])}
	\label{eq:LoS}
\end{align}
where $a$ and $b$ are the parameters related with environments,
and $x$ is the link distance between the transmitter and the receiver.
%$w=h_k/x$ is the ratio of a distance $x$ and a height between the transmitter and the receiver $\hij$.
\blue{
In real environment, \acp{UAV} can act as obstacles, e.g., \ac{UAV} in 1-layer can block the channel between ground and 2-layer \ac{AN}. 
Here, we assume existence of \ac{UAV} does not affect the air-to-ground channel since the density of \ac{AN} is low, therefore, the obstruction caused by \ac{UAV} is negligible.
}
From \eqref{eq:LoS}, we can see that the \ac{LoS} probability increases with $x$
which means higher altitude gives higher \ac{LoS} probability since there will be fewer obstructions.
The \ac{NLoS} probability is then given as \red{$\rho_{k}\coN(x)=1-\rho_{k}\coL(x)$.}

Since each link between a transmitter and a receiver can be in either \ac{LoS} or \ac{NLoS} with the probabilities,
$\rho_{k}\coL(x)$ and $\rho_{k}\coN(x)$, respectively,
we can divide a set of the $k$-layer transmitters into the ones in \ac{LoS} and the ones in \ac{NLoS} as
$\Phi_{k}\coL$ and $\Phi_{k}\coN$, respectively, which are non-homogeneous \acp{PPP}.
The densities of nodes in $\Phi_{k}\coL$ and $\Phi_{k}\coN$ with the distance $x$ from a receiver are, respectively, defined according to the link distance $x$ as $\lambda_{k}\coL(x)=2 \pi x \lambda_k \rho_{k}\coL\left(x\right)$ and $\lambda_{k}\coN(x)=2 \pi x \lambda_k -\lambda_{k}\coL(x)$.

%Note that hhe \ac{PPP} represent $j$-layer transmitter is seperated with the \ac{LoS} and \ac{NLoS} probability that gives non-homogeneous \ac{PPP}.
%Hence, use $\Phi_{j}\coL$ and $\Phi_{j}\coN$ to represent the transmitter that involved in $j$-layer and communication channel is \ac{LoS} and \ac{NLoS}, respectively.
%Furthermore, we can get the intensity function of non-homogeneous \ac{PPP} given by
%$\lambda_{ij}\coL(x)=2 \pi x \lambda_j \rho_{ij}\coL\left(x\right)$ and $\lambda_{ij}\coN(x)=\lambda_j-\lambda_{ij}\coL(x)$.
%%Note that we can replace the \ac{LoS} channel probability replaced with other models.

We also consider different channel models for links in \ac{LoS} and \ac{NLoS}.
%We assume the \ac{LoS} and the \ac{NLoS} channel have different pathloss coefficient and the fading.
The pathloss exponents for \ac{LoS} and the \ac{NLoS} links are denoted by $\alpha\coL$ and $\alpha\coN$, respectively, and generally, $2\leq \alpha^{(\text{L})} \leq\alpha^{(\text{N})} \leq 6$.
%In general, $2\leq \alpha^{(\text{L})} \leq\alpha^{(\text{N})} \leq 4$.
We consider the Nakagami-$m$ fading for the channels of \ac{LoS} and the \ac{NLoS} links,
of which channel gains are respectively presented by
$G\coL \sim \Gamma(m\coL,\frac{1}{m\coL})$ and $G\coN \sim \Gamma(m\coN,\frac{1}{m\coN})$.
Here, we use $m\coN=1$, which gives Rayleigh fading, i.e., $G\coN\sim\exp(1)$,
while $m\coL \geq 1$.
%Contrary, we use $m\coL \geq 1$ to represent more deterministic property of \ac{LoS} channel.
%which is used for the \ac{NLoS} links, and bigger $m\Co$ can be used for \ac{LoS} channels which are more deterministic channel due to the small variance.

%\subsection{Association Rules}/
\subsection{Association Rule}

In this paper, we assume a receiver connects to the transmitter,
which has the strongest average received power described in \cite{ChoLiuLee:18}. This can be applied to the scenario that in the presence of \ac{UAV} based \acp{BS} \cite{ZenZhaLim:16},
%\blue{[JL: Add one or two references that use a UAV as a BS]}
a user selects a \ac{BS} to receive its data. Based on the association rule, we can present the selected transmitter's coordinates $\bold{x}_\text{main}$ as
%
%x
\begin{align}\label{eq:S_SN_rule}
	\bold{x}_\text{main}
	&=
		\underset{\bold{x}\in\Phi_k, k\in\mathcal{K}}{\operatorname{arg\,max}} P_k \normx ^{-\alpha_\bx}
\end{align}
where $\bx_\text{rec}$ is the coordinates the receiver is located, and
$\alpha_\bx$ is the pathloss exponent of the link between the transmitter at $\bold{x}$ and the receiver.
%We denote the pathloss of the transmitter with $\alpha_\bx$.

% The association scenario is rule of communication between transmitter and receiver which affects \ac{PDF} of main link distance and interference regions that related with the \ac{STP}.
% We consider the receiver connect the transmitter considering pathloss and distance.
% Hence, the main link transmitter is represented by
%\begin{align}
%\bold{x}_\text{main}&=\underset{\bold{x}\in\Phi_k, k\in\mathcal{K}}{\operatorname{arg\,max}} \normx ^{-\alpha_\bx} \label{eq:S_SN_rule}
%\end{align}
%where $\bx$ is used for location of the transmitter and $\bx_\text{rec}$ is used for the receiver.
%We denote the pathloss of the transmitter with $\alpha_\bx$.
%Since we assume the same power across the \ac{UAV} we call the association scenario as strongest transmitter connection.
%However, by \cite{SinDhiAnd:13} and our approach, strongest transmitter connection considering different power is easily derived.

Based on the association rule above, we can determine the \ac{PDF} of the distance for \emph{the main link} from a selected transmitter to a receiver.
In conventional terrestrial networks, the \ac{PDF} of main link distance is
determined by the transmission power, the pathloss exponent, and the link distance.
However, in \acp{AN}, we need to additionally consider the \ac{LoS}/\ac{NLoS} probabilities for all links to the transmitters.  We denote the channel environment by \red{$c \in \{\text{N},\text{L}\}$}, where $c= \text{N}$ and $c= \text{L}$, respectively, means the LoS and NLoS environments of the link.
The \ac{PDF} of main link distance in $c$ channel condition is presented in following lemma.
\begin{lemma}\label{L:PDFM}
When a transmitter in the $j$-layer under the channel environment $c$ is selected,
the \ac{PDF} of main link distance $Y_{j}\!\Co$ is given by
%The \ac{PDF} of the main link distance when the main link transmitter is in the $j$-layer and the channel is $(C)$ is given by
%
%
\begin{align}
	&f_{Y\ijCo} \!(y)
	\!=\!
		\frac{f_{V\ijCo}(y)}{\mathcal{A}\ijCo}
		\!\prod_{
			\substack{
				k\in\mathcal{K}, \cb\! \in \!\{\!\text{L,N}\!\}\!,
				\\(k, \cb)\neq(j,c)
			}
		}\!
		\bar{F}_{V_{k}\To}\!\left( \!\left(\! \frac{P_k y^{\alpha\Co}}{P_j}\!\right)^{\!\frac{1}{\alpha\To}}\!\right)\!\!\label{eq:lem_1}
\end{align}
%
%
%where $\hat P_{j,k}=P_k/P_j$ is the power ratio of transmitters in the $k$-layer and the $j$-layer,
%and $\hat\alpha\To=\alpha\To/\alpha\Co$ is ratio of the pathloss.
where $\mathcal{A}\ijCo$ is association probability given by
\begin{align}
\mathcal{A}\ijCo	\!=\! \int\displaylimits_{y>0}
f_{V\ijCo} \!(y)\!
\!\prod_{
	\substack{
		k\in\mathcal{K}, \cb\! \in \!\{\text{L,N}\},
		\\(k, \cb)\neq(j,c)
	}
}\!
\bar{F}_{V_{k}\To}\!\left( \!\left(\! \frac{P_k y^{\alpha\Co}}{P_j}\!\right)^{\!\frac{1}{\alpha\To}}\!\right)\!. dy \label{eq:Asso}
\end{align}
%where $V_{k}\!\To$ is the distance to the nearest node among the nodes in the $k$-layer under the channel environment $\cb \in \{\text{N,L}\}$, and
%$ \bar{F}_{V\ijTo}(v)$ and $f_{V\ijTo}(v)$ are the \ac{CCDF} and the \ac{PDF} of
%%$F_X(x)$ is
%$V\ijTo$, given by %the \ac{CDF} of the random variable $X$.
%\begin{align}
%\bar{F}_{\!V\ijTo\!}(v)&\!=\exp\left[-\! \int_{\hij}^{\max(v,\hij)}\! 2\pi\lambda_j x \rho\ijTo\! \left(x\right)\!dx \right] \label{eq:S_CDF1} \\
%f_{\!V\ijTo\!}(v)&= 2\pi\lambda_jv \rho\ijTo \!\left(v\right)\exp\left[-\! \int_{\hij}^{v} 2\pi\lambda_j x \rho\ijTo \!\left(x\right)\!dx \right] \label{eq:S_PDF1}
%\end{align}
%where $f_{V\ijTo}$ is $0$ when $v<\hij$.
\red{
Here, $V_{j}\Co$ is the distance to the nearest node among the nodes in the $k$-layer under the channel environment $c$.
$\bar{F}_{V\ijCo}(v)$ and $f_{V\ijCo}(v)$ are the \ac{CCDF} and the \ac{PDF} of
$V\ijCo$, given by %the \ac{CDF} of the random variable $X$.
\begin{align}
 \bar{F}_{V\ijCo}(v)&=\exp\left[- \int_{\hij}^{\max(v,\hij)}\! 2\pi\lambda_j x \rho\ijCo \left(x\right)\!dx \right], \label{eq:S_CDF1} \\
 f_{V\ijCo}(v)&= 2\pi\lambda_jv \rho\ijCo \left(v\right)\exp\left[- \int_{\hij}^{v} 2\pi\lambda_j x \rho\ijCo \left(x\right)\!dx \right].\nonumber
\end{align}
The \ac{PDF} $f_{V\ijCo}$ is $0$ when $v<\hij$.
}
\end{lemma}
\begin{IEEEproof}
%The intensity function of non-homogeneous \ac{PPP} is given as
%
%\begin{align}
%\lambda\ijTo(x)=2 \pi x \lambda_j \rho\ijTo\left(x\right)
%\label{eq:A_PDF_1}
%\end{align}
%which gives \ac{CDF} of nearest node distance as
%
%\begin{align}

The \ac{CDF} of $V\ijCo$ is given by
\begin{align}
F_{V\ijCo}(v)
%& =
%\mathbb{P}\left( V\ijCo \leq v\right)
%\overset{}{=}
%1-\mathbb{P}\left(|\Phi_j\Co\cap B(o, v)|=0\right) \nonumber \\
&
\overset{(a)}{=}
1-\exp\left[-\int_{\hij}^{\max (v, \hij)} \lambda\ijCo(x) dx\right]\label{eq:A_PDF_2}
\end{align}
%The \ac{CDF} of $V\ijTo$ is given by
%\begin{align}
%	F_{V\ijTo}(v)
%	& =
%		\mathbb{P}\left( V\ijTo \leq v\right)
%   \overset{}{=}
%		1-\mathbb{P}\left(|\Phi_j\To\cap B(o, v)|=0\right) \nonumber \\
%	&
%	\overset{(a)}{=}
%		1-\exp\left[-\int_{\hij}^{\max (v, \hij)} \lambda\ijTo(x) dx\right]\label{eq:A_PDF_2}
%\end{align}
%
%Since nearest distance is smaller than $v$ equivalent to there is at least one transmitter, (a) is given.
%Addition,
\red{where (a) is from the void probability of \ac{PPP}, and from \eqref{eq:A_PDF_2}, we have \eqref{eq:S_CDF1}.}% and differentiating \eqref{eq:A_PDF_2} with respect to $v$ gives \eqref{eq:S_PDF1}.
%
%
%Since the main link have smallest pathloss, by considering the channel and the nearest distance, the \ac{PDF} of main link distance derived as
%Since the main link have smallest pathloss, by considering the channel and the nearest distance, probability that $j$-layer under channel $c$ is selected with distance bigger than $y$ is given by
Since the main link have smallest pathloss, probability that main link distance is smaller than $y$ when $\bx_\text{main}\in\Phi\ijCo$ is given by
\begin{align}
%&\mathbb{P}\left( Y\ijTo \leq y \;\cap\; \text{main link in }j\text{-layer under channel } c \right)  \nonumber \\
&\red{\mathbb{P}\left( Y\ijCo \leq y \;\cap\; \bx_\text{main}\in\Phi\!\ijCo \right) } \nonumber \\
&=\int_{0}^{y} f_{V\ijCo}(v) \mathbb{P}\left(\bx_\text{main}\in\Phi_j\Co \;\cap\; V\ijCo=v \right) d v \label{eq:A_PDF_5} \\
&\overset{(a)}{=}\!\int_0^{y} \!f_{V\ijCo}(v)\! \prod_{\substack{k\!\in\!\mathcal{K}\!, \cb\! \in\! \{\!\text{L,N}\!\}\!,\\(k, \cb)\neq(j,c)}}\!\mathbb{P}\!\left[  P_j v\aCo \!\geq\! P_k \!\left(V\ikTo\right)\!\aTo \!   \right]\! dv \nonumber
\end{align}
where (a) from \eqref{eq:S_SN_rule}.
Therefore, we derived the association probability as \eqref{eq:Asso} by $y\to\infty$.
Furthermore, we can derived the \ac{PDF} of the main link distance as \eqref{eq:A_PDF_2}.
% Furthermore, we can derived the \ac{PDF} of the main link distance as \eqref{eq:A_PDF_2}.
%Since the result is the product of \ac{PDF} and \ac{CDF}, we can derive \eqref{eq:lem_1}.
\end{IEEEproof}

\section{Interference Analysis and Successfully Transmission Probability}

%The distance between the transmitter and the receiver is denoted by $x=\sqrt{h_{ij}^2+r^2}$, where $r$ is the horizontal distance between a transmitter and a receiver and $h_{ij}$ is the altitude difference between layers.

In this section, we analyze the Laplace transform of the interference considering association rules. Then, we derive the \ac{STP} of the \ac{MAN} and the upper bound of the density of \ac{AN} that maximize the \ac{STP} of the \ac{MAN}.

\subsection{Laplace Transform of the Interference}

%In the \ac{MAN},
%when the $j$-layer transmitter under the channel environment $c$ is selected,
%%when the main link is under the channel environment $c$ and the link distance is $d$,
%the interference from transmitters in the $k$-layer under the channel environment $\cb$ is given by
\red{
In the \ac{MAN}, we first consider interference from specific layer and channel environment. Since there is no interferer which have stronger power than main link transmitter, we analyze interference from specific layer and channel to analyze total interference.
Here, the interference from transmitters in the $k$-layer under the channel environment $\cb$ is given by
}
\begin{align}
	I\ikTo&=\sum_{\bx\in\Phi\ikTo } P\ikTo (\normx ) \label{eq:Iij}
  \vspace{-3mm}
\end{align}
where $P_k\!\To$ is the received power from a transmitter which is given by
\begin{align}
P\ikTo(x)=P_k G\To x^{-\alpha\To}.
	\label{eq:Pij}
\end{align}
\red{Here, we represent distance between the transmitter and the receiver as $x=\normx$.}
The Laplace transform of the interference is given in the following lemma.
\red{In the lemma, we use $\jcy$ to represent the main link under channel environment $c$ with distance $y$ is selected.}

\begin{lemma}\label{L:LAPI}
When a transmitter in the $j$-layer under the channel environment $c$ with distance $y$ is selected, the Laplace transform of the interference from the transmitting nodes in the $k$-layer under the channel environment $\cb$ is given by \eqref{eq:IntLap}, which presented on the top of next page,
\begin{figure*}[!tbp]\vspace*{1pt}\begin{align}
&\Lap{I\ikTo | \jcy}(s)=\exp\left[- 2\pi \lambda_k \left\{ \int_{\max \left(\RI(y), \hik\right)}^\infty  x\rho\ikTo\left(x	\right)\left(1-\left(\frac{1}{1+\frac{sP_kx^{-\alpha\To}}{m\To}}\right)^{m\To} \right) dx  \right\}  \right]	\label{eq:IntLap}
\end{align}\hrulefill\normalsize\end{figure*}%
%where $\RI(y)=\left(   P_k y^{\alpha\Co}/P_j \right)\!^{1/\alpha\To}$.
\red{where $\RI(y)$ is% represented as
\begin{align}
\RI(y)=\left(   P_k y^{\alpha\Co}/P_j \right)\!^{1/\alpha\To}. \label{eq:Int_free_zone}
\end{align}}
% and $\jcd$ represent that a transmitter in the $j$-layer under the channel environment $c$ is selected with distance $d$.
\end{lemma}
\begin{IEEEproof}
The Laplace transform of the interference is
\begin{align}
&\Lap{I\ikTo | \jcy}(s) \nonumber \\
%&=\mathbb{E}_{\Phi\ikTo,G\To} \left[ \prod_{\bx\in\Phi\ikTo}	\exp\left(-sP_k G\To x\aCo		\right)		 \;\biggr|\; \jcy  \right] \nonumber \\
&\overset{(a)}{=}\mathbb{E}_{\Phi\ikTo} \left[\!\prod_{\bx\in\Phi\ikTo }\!	\left(	\!\frac{1}{1+\frac{sP_k x^{-\alpha\To}}{m\To}}\!\right)^{m\To} \biggr|\;\jcy\right]		\label{eq:Frint}
\end{align}
%
%where $d$ is the main link distance and
where (a) is from the expectation over channel $G\To$ which gives the \ac{MGF} of Gamma distribution as \cite{ChoLiuLee:18}.
Since $\lambda_{k}\To(x)=2 \pi x \lambda_k \rho_{k}\To\left(x\right)$,
%From the intensity function of the non-homogeneous \ac{PPP},
the \ac{PGFL} of non-homogeneous \ac{PPP} needs to be obtained as\cite{HaeGan:08}
\begin{align}
&\mathbb{E}   \left[ \prod_{\bx\in\Phi\ikTo} f(\bx)\; \biggr|\; \jcy \right] \nonumber \\
 &=\exp\left(-2\pi\lambda_k \int_{R'}^\infty x(1-f(x))\rho\ikTo\left(x\right)dx\right). \label{eq:IL_PGFL}
\end{align}
In \eqref{eq:IL_PGFL},
%$R'$ is used for the situation that interferer is can not exist closer than $R'$, e.g., association rules and interference cancellation guarantee no interferer in the area.
selecting a transmitter in the $j$-layer under the channel environment $c$ by \eqref{eq:S_SN_rule}
means there is no interfering node in the $k$-layer under channel environment $\cb$, closer than $R'=\max\left(\RI(y),\;\hik\right)$.
Combined with \eqref{eq:Frint} and \eqref{eq:IL_PGFL}, we obtain the Laplace transform interference under condition $\jcy$ as \eqref{eq:IntLap}
%We use $R'=\max(\RI(d),\;\hik)$ since distance of interferer should be bigger than both height $\hik$ and interference free region $\RI(d)$.
%Finally, using \eqref{eq:IL_PGFL}, we derive the Laplace transform of the interference.
\end{IEEEproof}

%  \begin{figure}[t!]
%     \begin{center}
%    {
% 	 \psfrag{AAAAAAAAAAA}[][][0.7]{$s=1*10^3$}
% 	 \psfrag{AAAAAAAAAAB}[][][0.7]{$s=5*10^3$}
% 	 \psfrag{AAAAAAAAAAC}[][][0.7]{$s=1*10^4$}
% 	 \psfrag{AAAAAAAAAAD}[][][0.7]{$s=5*10^4$}
% 	 %\psfrag{AAAAAAAAAAE}[][][0.7]{$h_{ij}$ (m)}
% 	 %\psfrag{AAAAAAAAAAF}[][][0.7]{$\Lap{I_{ij}}^{(\mu)}(s)$}
% 	 \psfrag{AAAAAAAAAAE}[][][0.7]{$h_{1}$ (m)}
% 	 \psfrag{AAAAAAAAAAF}[][][0.7]{$\Lap{I_1}(s)$}
% 	 \includegraphics[width=1.00\columnwidth]{Figures/Fig_Lap_h.eps}
%     }
%     \end{center}
%     \caption{
% 		The Laplace transform of interference for scenario $(\mu)$ according to the altitude of \ac{AN} which act as interference.
%     Density of interferer $\lambda_j=10^{-5}$ and parameters used in this figure are explained and specified in the next section.
% 		%Lines and asterisks are analytical and simulation results respectively, and $\sigma^2=0$ and $\sigma^2= 10^{-6} P$ cases are presented for each horizontal distance.
% 		% Vertical lines separate tendency of \ac{STP} for $r=40$.
% 		 }   \label{fig:F_Lap}
% \end{figure}
\red{
 From Lemma~\ref{L:LAPI} and property of the Laplace transform,
}
 we can obtain the Laplace transform of the sum of the interference and noise as
\begin{align}
\Lap{\mathcal{I} | \jcy}(s)&\!=\!	\exp(-s\sigma^2)\!\prod_{k\in\mathcal{K},\Tin}\!\Lap{I\ikTo | \jcy}(s)	\label{eq:prodL}
\end{align}
where $\mathcal{I}=\sum_{k\in\mathcal{K}}^{\Tin} I\ikTo+\sigma^2$ and $\sigma^2$ is for the noise power.

\subsection{Successful Transmission Probability}

\red{In this subsection, we define the \ac{STP} when the distance and the channel environment is given. Then, we derive the \ac{STP} of the \ac{MAN} by using association probability and the \ac{PDF} of the main link distance.}
When the main link is in the channel environment $c$ with the link distance $y$,
the \ac{STP} is defined using \ac{SINR} as
%We analyze the probability of satisfying the required target \ac{SINR}, i.e., \ac{STP}, defined as
\begin{align}
p_{j}\Co(y)&=\mathbb{P}\left[\text{SINR}_{j}\Co(d)	>\beta	\right]			\label{eq:STPijd}
\end{align}
where $\text{SINR}_{j}\Co(d)=P\ijCo(d)/\mathcal{I}$, and $\beta$ is the target \ac{SINR}, which is related with the transmission rate. When the association rule in \eqref{eq:S_SN_rule} is used, the \ac{STP} of \ac{MAN} is presented in the following lemma.
% We assume $\beta$ is same for every receiver and transmitter pairs.
% Second, we derive \ac{STP} /when distance is fixed and channel is known.
% Finally, \ac{STP} of nearest receiver connection \ac{ASE} can be derived with channel probabilities and \ac{PDF} of nearest receiver distance.

%
\begin{lemma}\label{L:STPF}
The \ac{STP} of the \ac{MAN} is given by
%When a transmitter in the $j$-layer under the channel environment $c$ is selected using \eqref{eq:S_SN_rule},
% the \ac{STP} is given by
%The probability that a transmission in the $j$-layer under the channel environment $c$ is selected, and the communication succeed is given by
\begin{align}\label{eq:meanSTP}
	\mathcal{P}
	=
		\sum_{
				j \in \mathcal{K}, \,
				c \in \{\text{L,N}\}
				%(j,C)\in\mathcal{K}\times\{\text{L,N}\}
			}
			%\mathcal{P}_{j}\Co.
 	 		%\mathcal{P}_{j}\Co=
		 \int_{h_j}^\infty p\ijCo(y) f_{Y\ijCo}(y) \mathcal{A}\ijCo dy
%		 \int_{h_j}^\infty p\ijCo(y) f_{Y\ijCo}(y) dy
\end{align}
where
$f_{Y\ijCo}(y)$ and $\mathcal{A}\ijCo$ is in \eqref{eq:lem_1} and \eqref{eq:Asso},
$\pijc(y)$ is % the \ac{STP} when the main link distance is $d$ with channel $(C)$ that given by
\begin{align}
&\pijc(y)
=\left. \sum_{n=0}^{m\Co -1}\frac{(-s)^n}{n!}\frac{d^n}{d s^n}\Lap{\mathcal{I}| \jcy}  \left( s	\right) \right|_{s=\mathcal{S}_j\Co(y)},  \label{eq:STPF}\\
&\mathcal{S}_{j}\Co(y)=\frac{m\Co\beta}{P_jy^{-\alpha\Co}}, \label{eq:S_j}
\end{align}
%
%
%$\mathcal{S}_{j}\Co(y)=\frac{m\Co\beta}{P_jd^{-\alpha\Co}}$,
and $\Lap{\mathcal{I}| \jcy}  \left( s	\right)$ is in \eqref{eq:prodL}.%\blue{[JL:$\leftarrow$ add the equation number]}
%\begin{align}
%\mathcal{S}_{j}\Co(d)=\frac{m\Co\beta}{P_jd^{-\alpha\Co}} \label{eq:Sj}
%\end{align}
%
\end{lemma}
\begin{IEEEproof}
From the definition of \ac{STP}, % in \eqref{eq:STPijd},
we have the conditional \ac{STP} when a transmitter in $j$-layer under channel environment $c$ is selected with the main link distance $j$ represented as
%\blue{[JL:Add how you obtain (15)--- (XX)]}
%
%
\begin{align}
&\pijc(y)%=\mathbb{P}\left[ G\Co 	> \frac{\beta \mathcal{I}}{P_jy^{-\alpha\Co}}	\;\biggr|\; \jcy\right]\nonumber \\
\overset{(a)}{=}\mathbb{E} \left[	1-\frac{1}{\Gamma(m\Co)}\gamma\left(	m\Co, \frac{m\Co\beta}{P_jy^{-\alpha\Co}}	\mathcal{I} 	\right) \;\biggr|\; \jcy			\right]  \nonumber \\
&\overset{(b)}{=}\mathbb{E}\left. \left[\sum_{n=0}^{m\Co -1} \frac{\left(s \mathcal{I} \right)^n}{n!} \exp\left(- s \mathcal{I}	\right) \;\biggr|\;	\jcy	\right]	\right|_{s=\mathcal{S}_j\Co(y)}		\label{eq:ProofSTP}
\end{align}
where (a) follows from the Gamma distribution of channel gain and (b) follows from the property of lower incomplete Gamma function.
Notice that we derived \eqref{eq:S_j} from (b).
Using the property of the following Laplace transform%\blue{[JL: Add a reference]}, %following equality holds
\begin{align}
(-\mathcal{I})^n \Lap{\mathcal{I}}(s)=\frac{d^n}{d s^n}\Lap{\mathcal{I}}(s)
\end{align}
we obtain \eqref{eq:STPF}.
Furthermore, from the \ac{PDF} and association probability in the Lemma~\ref{L:PDFM}, we obtain \eqref{eq:meanSTP}.
\end{IEEEproof}

%From Lemma~\ref{L:STPF}, the total \ac{STP} is derived as
%\begin{align}
%\mathcal{P}=\sum_{(j,C)\in\mathcal{K}\times\{\text{L,N}\}} \mathcal{P}_{j}\Co. \label{eq:E_totalSTP}
%\end{align}
\red{When \acp{UAV} are deployed as \acp{BS} which serve for ground receiver, it is important to maximize \ac{STP}. In our works, we analyze optimal density of the transmitter.}
As shown in \eqref{eq:meanSTP}, it is \red{hard to present} \ac{STP} in a closed form, \red{hence}, hard to obtain the optimal densities of each \ac{AN} layer that maximize \ac{STP}.
However, in the following corollary, we present the closed-form upper bound of the optimal densities for a special case.

%Since we can not derive the optimal density of each \ac{AN} because the total \ac{STP} is not closed form, we get the upper bound for the optimal density.
\begin{corollary}\label{cor:UpperBound}
For the case of $m\coN=m\coL=1$, when the optimal density of the $j$-layer \ac{AN} is $\lambda_j^*$, its upper bound is
\begin{align}
	\lambda_j^* \leq \lambda_j^b  = \frac{1}{2\pi  \epsilon_j(\mathcal{S}_j\coL(\hij)) } \label{eq:L_Up_1}
\end{align}
where $\mathcal{S}_j\coL(\hij)$ is in \eqref{eq:S_j}, and
%the upper bound of the optimal density satisfying $\lambda_j^*\leq \lambda_j^b$ and
$\epsilon_j(s)$ is given by
\begin{align}
&\epsilon_j(s)  = \int_{\hij}^\infty x\left(1-\frac{\rho\ijcoL (x)}{1+sP_j x\acoL}-\frac{\rho\ijcoN(x)}{1+sP_j x\acoN}  \right)dx. \label{eq:L_Up_1}
\end{align}
\end{corollary}
\begin{IEEEproof}
See Appendix~\ref{app:bound}.
\end{IEEEproof}

\red{
In Corollary~\ref{cor:UpperBound},
the upper bound $\lambda_j^b$ is
only affected by the network parameter of the $j$-layer \ac{AN} such as $h_j$,
but independent
with the density, height, and transmission power of other \acp{AN}.
Hence, the upper bound of each layer's density in \ac{MAN} can be determined independently each other.
Although function $\epsilon_j(s)$ is not in the closed form, it is easy to evaluate and analyze.}
\footnote{\red{Note that the upper bound of optimal density $\lambda_j^b$ is decrease with height $\hij$ under condition $\beta\hij^{\alpha\coL}>1$, $\hij>1$ which is omitted in this paper. The conditions are readily achieveable for aerial networks.}}
%  each layer \ac{AN}
%i.e., when we design the \ac{MAN} which float in given heights, we can find the upper bound of the density of each \ac{AN}.
%
Furthermore, due to the above independency,
we can also obtain the upper bound of the total density of \ac{MAN} as
$\lambda_T^b=\sum_{k\in\mathcal{K}}\lambda_k^b$.

\section{Numerical Results}
\begin{figure}[t!]
	\begin{center}
		{
			\psfrag{AAAAAAAAAAA}[][][0.9]{$m\coL=1$}
			\psfrag{AAAAAAAAAAB}[][][0.9]{$m\coL=3$}
			\psfrag{AAAAAAAAAAC}[][][0.9]{$m\coL=20$}
			\psfrag{AAAAAAAAAAD}[][][0.9]{$m\coL=100$}
%			\psfrag{AAAAAAAA}[][0.9]{$\rho_{j}\coL(x)=1/(1+a e^(-b))$}
			\psfrag{AAAAAAAC}[][0.9]{$\rho_{j}\coL(x)=1$}
			\psfrag{EAAAAAAAAAA}[][][0.8]{$h_1$ (m)}
			\psfrag{FAAAAAAAAAA}[][][0.8]{Successful Transmission Probability}
			\includegraphics[width=1.00\columnwidth]{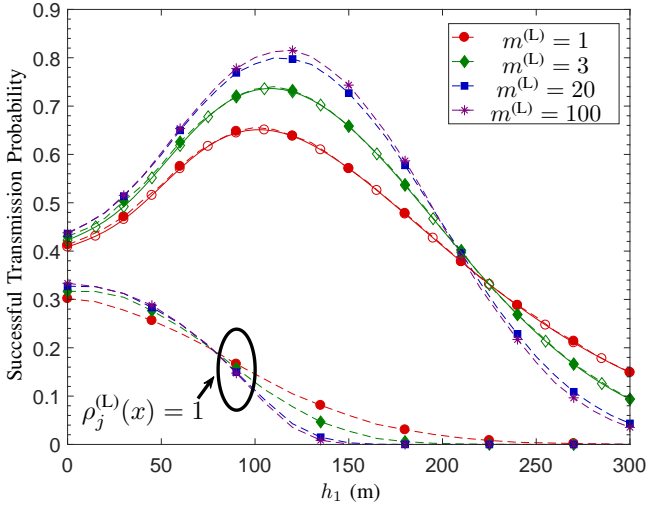}
      \vspace{-8mm}
		}
	\end{center}
	\caption{
		\ac{STP} of single \ac{AN} according to the altitude $h_1$ and the \ac{LoS} coefficient $m\coL$ when $\lambda_1=10^{-5}$. $\rho_{j}\coL(x)=1$ represent the \ac{STP} when every link is under \ac{LoS} channel.
      \vspace{-6mm}
		%Due to the distance-\ac{LoS} probability tradeoff that depends on height, optimal height exists.
	}   \label{fig:F_O_One}
\end{figure}

In this section,
we present the numerical results to evaluate our analysis on the \ac{STP} of \ac{MAN} with single or two layers of \acp{AN} under the interference-limited environments. i.e., $\sigma^2=0$.
	For the numerical results, we use %the \ac{SINR} threshold
	$\beta=0.7$, $P_k=1$ for all $k$,
	$\alpha\coL=2.5$, and $\alpha\coN=3.5$.
	Except for Fig.~\ref{fig:F_O_One}, $m\coN=1$ and $m\coL=1$ are used.
	Moreover, we use $a=12.4231$ and $b=0.1202$ are used for the \ac{LoS} probability, which are determined for the urban area environment in \cite{AlKanLar:14}.% with \ac{LoS} parameters $a=12.4231$ and $b=0.1202$.
	%The transmission power is $P_k=1$ for all $k$ and the pathloss exponents for \ac{LoS} and \ac{NLoS} channel are $\alpha\coL=2.5$ and $\alpha\coN=3.5$, respectively.
	% since the transmission power of \ac{UAV} experiences severe attenuation due to the long link distance.
% we evaluate the \ac{STP} of the \ac{MAN} that compsed of single or 2-layers of \acp{AN} considering interference limited environment.
%We use the threshold $\beta=0.7$ since power from \ac{UAV} experience severe attenuation due to the distance.
%Also, except Fig.~\ref{fig:F_O_One}, we consider $m\coN=1$, and $m\coL=1$ for tractable computation.%or tractable computation
%
%
%The value of $m^{(N)}$ is not fixed to $1$ in Fig.2. Need some modification in arrangement...
%
%Without loss of generality, throughout this section, we consider that the transmission power is $P_k=1$ for all $k$ and the pathloss exponents for \ac{LoS} and \ac{NLoS} channel are $\alpha\coL=2.5$ and $\alpha\coN=3.5$, respectively. Moreover, we consider the urban area environment in \cite{AlKanLar:14} with \ac{LoS} parameters $a=12.4231$ and $b=0.1202$.
%The pathloss coefficient is $\alpha\coL=2.5$, $\alpha\coN=3.5$, and power is $P_k=1$, for all $k$, and the parameters for \ac{LoS} are set to $a=12.4231$ and $b=0.1202$, which derived for the urban area in \cite{AlKanLar:14}.

\red{Fig.~\ref{fig:F_O_One} shows the \ac{STP} as a function of the altitude $h_1$ in single \ac{AN}
for different values of channel coefficient $m\coL=\{1,3,20,100\}$,
 where $\lambda_1=10^{-5}$[nodes/$m^2$]. We shows the \ac{STP} when \ac{LoS} probability is $1$ which represented with $\rho_{j}\coL(x)=1$.}
	%Here, we use $m\coL=\{1,3,20,100\}$ to verify the impact of \ac{LoS} channel on the \ac{STP}.
	Simulation results, obtained from Monte Carlo simulations, are presented by the dashed lines with filled markers, while analysis results for $m\coL=\{1,3\}$ are presented by the solid lines with unfilled markers.
	From Fig.~\ref{fig:F_O_One}, we can first see that the simulation results match well with the analysis.
	% when $m\coL=\{1,3\}$
	%%
	We can also see that for all $m\coL$,
	the \ac{STP} first increases and then decreases with $h_1$.
	For small $h_1$, as the height increases, the \ac{LoS} channel probability increases, which makes the main link power stronger and results in the higher \ac{STP}. However, as $h_1$ keeps increasing, the main link distance also increases, which makes the main link power smaller.
	As a result, the optimal value of height can be obtained from the tradeoff between the link distance the the \ac{LoS} probability.
\red{When the \ac{LoS} probability is $1$, there is no tradeoff, the \ac{STP} of the \ac{AN} decrease with height, which is same for $\rho_j\coL(x)=0$.}

\red{
	Furthermore, by comparing the results for different $m\coL$, we can see that for small $h_1$, larger \ac{LoS} coefficient $m\coL$ give higher \ac{STP}.
	In smaller $h_1$, power from the main link is dominant and mostly \ac{LoS} channel. Since larger \ac{LoS} channel coefficient gives smaller variance to the main link channel in smaller $h_1$, larger $m\coL$ can give higher \ac{STP}.
	Contrary, in larger $h_1$, power from the interference is dominant. Therefore, larger \ac{LoS} channel coefficient gives smaller variance to the interference channel which gives lower \ac{STP}.
	However, trends of \ac{STP} according to the height are the same for different $m\coL$.
	Therefore, we use $m\coL=1$ in the following numerical results because it can give sufficient insights on the performance of \ac{MAN}.
}
%Because the \ac{STP} of \ac{MAN} more affected by the pathloss than \ac{LoS} coefficients, where pathloss is affected by the height and the density which is related with the \ac{LoS} channel probability and number of interferers in \ac{AN}.
%Therefore, analysis based on $m\coL=1$ is reasonable because it is sufficient to gives insights of \ac{MAN}.

%Larger $m\coL$ reulsts in higher \ac{STP} when \ac{LoS} channel is dominent in main link and lower \ac{STP} when \ac{LoS} channel is dominent
%\purp{By comparing the results for different $m\coL$ in Fig.~\ref{fig:F_O_One}}, \purp{we can see that} trends are similar because they depends on the density and the altitude of \ac{AN}, not $m\coL$. \purp{The optimal value of height for different $m\coL$ are also similar because that the main links at the optimal height are mostly \ac{LoS} channel.}
%%Since \purp{the} \ac{LoS} channel probability depends on the heights of \purp{\acp{UAV}}, although $m\coL$ is different, optimal height is similar.
%However, as $m\coL$ becomes larger, the \ac{STP} increases because of more non-deterministic property of main link power that increase the \ac{STP}\purp{[CW:what is the non-deterministic property?]}.
%On the other hands, in high altitudes, larger $m\coL$ results in smaller \ac{STP} since \purp{the interferers experience the \ac{LoS} channel}.\purp{[CW:I can't understnad the correct meaning of this paragraph]}

\begin{figure}[t!]
	\begin{center}
		{
			\psfrag{EAAAAAAAAAA}[][][0.8]{$h_1$ (m)}
			\psfrag{FAAAAAAAAAA}[][][0.8]{$\lambda_1$ (number/$m^2$)}
			\includegraphics[width=1.00\columnwidth]{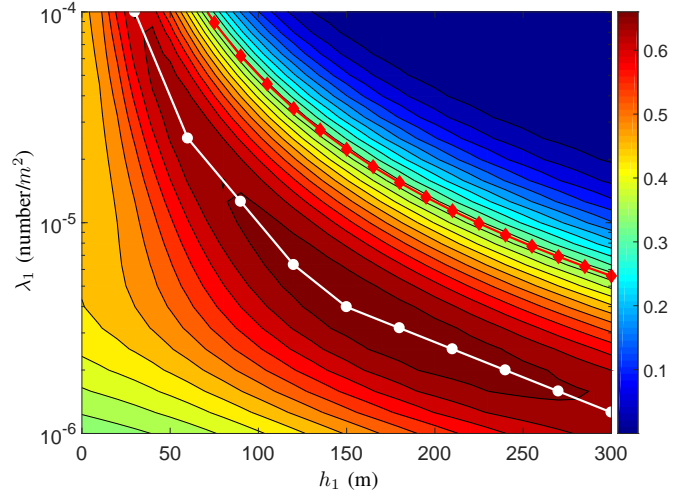}
      \vspace{-8mm}
		}
	\end{center}
	\caption{
		\ac{STP} of single \ac{AN} as functions of the density $\lambda_1$ and the height $h_1$ when $m\coL=1$.
			The white line marked by circles represents the optimal density when the height is given and the red line marked by diamonds represents the upper bound of the optimal density.
      \vspace{-6mm}
		%The \ac{STP} of sigle-\ac{AN} as function of the density $\lambda_1$ and the height $h_1$ when $m\coL=1$. A white line with circles represent optimal density in given height and a red line with diamonds represents the upper bound of optimal density.
	}   \label{fig:F_O_Two}
\end{figure}

Fig.~\ref{fig:F_O_Two} shows the contour of \ac{STP} as a function of the altitude $h_1$ and the density $\lambda_1$ of \ac{UAV} in single \ac{AN} where $m\coL=1$.
We represent the optimal density as a white line with circles and the upper bound of optimal density, obtained from Corollary~\ref{cor:UpperBound}, as a red line with diamonds.
By comparing the optimal density and the upper bound of the optimal density,
we can notice that the trends according to $h_1$ are the same.
From Fig.~\ref{fig:F_O_Two}, we can also see that as the height increases, the optimal density and its upper bound decrease. This is because the number of interfering links in \ac{LoS} increase with $h_1$, so the interference becomes larger.
%with the heights and trends of optimal density follows the upper bound.
%In higher altitude, larger density results in more \ac{LoS} channel interferer, hence optimal density decrease with heights.

\begin{figure}[t!]
	\begin{center}
		{
			\psfrag{EAAAAAAAAAA}[][][0.8]{$\lambda_1$ (number/$m^2$)}
			\psfrag{FAAAAAAAAAA}[][][0.8]{$\lambda_2$ (number/$m^2$)}
			\includegraphics[width=1.00\columnwidth]{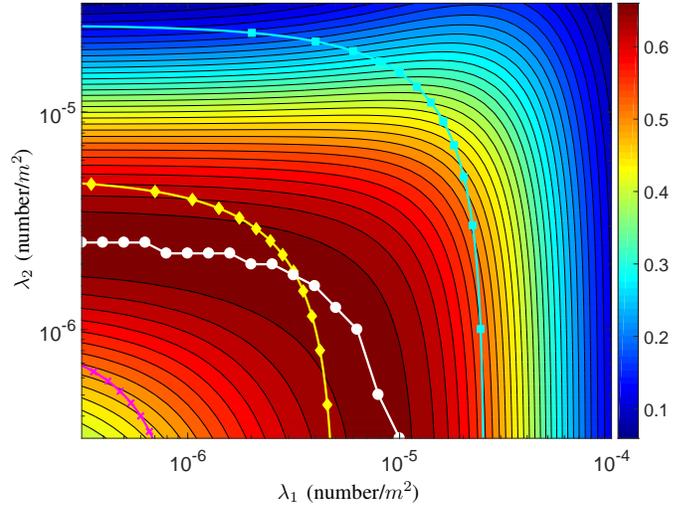}
      \vspace{-8mm}
		}
	\end{center}
	\caption{
		\ac{STP} of 2-layer \ac{MAN} as functions of the density of 1-layer $\lambda_1$ and the density of 2-layer $\lambda_2$ of \ac{UAV} when $m\coL=1$, $h_1=100$, and $h_2=200$.
		%as function of the density of $2$-layer $\lambda_2$ and the density of $1$-layer $\lambda_1$ of \ac{UAV} in a 2-layer \ac{MAN} where $m\coL=1$, $h_1=100$, $h_2=200$.
		The white line represents the optimal $\lambda_2$ in given $\lambda_1$.
    The magenta, yellow, and cyan lines with symbols represents the area which have same total density
    $\lambda_T=\lambda_1+\lambda_2=\{10^{-6}, 10^{-5.3}, 10^{-4.6}\}$
      \vspace{-0mm}
%    $\lambda_T=\lambda_1+\lambda_2$.
	}   \label{fig:F_O_Three}
\end{figure}

%\begin{figure}[t!]
%	\begin{center}
%	{
%		\psfrag{EAAAAAAAAAA}[][][0.8]{$\lambda_T$ (number/$m^2$)}
%		\psfrag{FAAAAAAAAAA}[][][0.8]{$\lambda_1/\lambda_T$}
%		\includegraphics[width=1.00\columnwidth]{Figures/NR_O_multi_LTR1.eps}
%	}
%\end{center}
%\caption{
%	Fig.~\ref{fig:F_O_Three} displays the \ac{STP} as function of the ratio of the $1$-layer $\lambda_1/\lambda_T$ and the total density $\lambda_T$ in the 2-layer \ac{MAN} where $m\coL=1$, $h_1=100$, $h_2=200$.
%	A white line represent optimal $\lambda_1/\lambda_T$ in given $\lambda_T$.
%}   \label{fig:F_O_Four}
%\ end{figure}

Fig.~\ref{fig:F_O_Three} shows the contour of \ac{STP} of \ac{MAN} with two \ac{AN}
as a function of the density of $1$-layer $\lambda_1$
and the density of $2$-layer $\lambda_2$ of \ac{UAV},
where $m\coL=1$, $h_1=100$, $h_2=200$.
In this figure, the optimal density of $2$-layer $\lambda_2$ is presented by a white line with circles.
%The magenta, yellow, and cyan line with crosses, diamonds, and squares represent the \acp{STP} of the \ac{MAN} that have the total density as $\lambda_T=\lambda_1+\lambda_2=\{10^{-6}, 10^{-5.3}, 10^{-4.6}\}$[nodes/$m^2$], respectively.
\red{
The lines with color and symbol represent the \acp{STP} of the \ac{MAN} that have the total density as $\lambda_T=\lambda_1+\lambda_2=\{10^{-6}, 10^{-5.3}, 10^{-4.6}\}$[nodes/$m^2$], respectively.
We can see when density of corresponding \ac{AN} is low, the \ac{STP} increase with the density, and when density is high, the \ac{STP} decrease with density which is same with the result of single-layer \ac{MAN}.
}
Furthermore, from the colored lines, we can get the relationship of optimal ratio of densities when the total density of \ac{MAN} is given.
\red{
When the total density of the \ac{MAN} is small, e.g., the magenta line,
the optimal density is $\lambda_2=\lambda_T$.
However, for the large total density, e.g., the cyan line, the optimal density is $\lambda_1=\lambda_T$. 
In other case, as shown in yellow line, we can see the optimal is neither $\lambda_1=\lambda_T$ nor $\lambda_2=\lambda_T$.
Since optimal height of larger \ac{AN} is low as shown in single \ac{AN}, optimal ratio of lower layer is increased with the total density. 
}

\section{Conclusion}
This paper models the \ac{MAN} which is wireless networks consist of multi-layers of \acp{UAV} that distributed in \ac{PPP} with different densities, heights, and powers.
We consider \ac{LoS} and \ac{NLoS} channel and the strongest transmitter association for the \ac{AN}.
Our approach is to derive the \ac{PDF} of the main link distance and the Laplace transform of the interference for the \ac{STP} analysis.
By analyzing the \ac{STP}, we show that each \ac{AN} in the \ac{MAN} have the upper bound of optimal density which is given by the function of the height of corresponding \ac{AN}.
In addition, our numerical results show the tradeoff caused by height of the \ac{AN}, affection of \ac{LoS} coefficient, significance of the upper bound of the optimal density, and optimal densities and optimal ratio of densities in the 2-layer \ac{MAN}.
Specially, our results show higher altitude \ac{AN} have sparser optimal density and show that the optimal ratio of densities in the 2-layer \ac{MAN} is changed with the total density of the \ac{MAN}.

\begin{appendix}\label{app:bound}
\subsection{Proof of Corollary~\ref{cor:UpperBound}}\label{app:UpperBound}
From Lemma~\ref{L:STPF},
the \ac{STP} can be represented by
\red{
\begin{align}
\mathcal{P}&=\sum_{j\in\mathcal{K}}\int_{\hij}^{\infty} \left(\varphi\ijcoL(y) + \varphi\ijcoN(y) \right)dy, \label{eq:A_bound_1}\\
%\end{align}
%
%
%where $\varphi\ijCo(y)$ is given by
%is product of \ac{STP} and \ac{PDF} in given channel and distance that represented as
%\begin{align}
%&  \varphi\ijCo(y)
%  \!=\!
%  	p\ijCo(y) f_{Y\ijCo}(y)
%	%\mathbb{P}\left[\bx_\text{main}\in \Phi\ijCo \cap \text{SINR}\ijCo(y)\geq \beta \cap d_\text{main}=y \right]
%	\nonumber\\
%  &\!=\!
%  2\pi\rho\ijCo \!\left(\!y\!\right)\! \lambda_j y  \exp \!\left[ \!  -2\pi\sum_{k\in\mathcal{K}}\lambda_k   \phi\ijkC\left(y, \mathcal{S}_j\Co(y)\right) \! \right]\label{eq:A_bound_2}
%\end{align}
%\red{
%\begin{align}
  \varphi\ijCo(y)
&=
2\pi\rho\ijCo \!\left(y\right)\! \lambda_j y  \exp \!\left[ \!  -2\pi\sum_{k\in\mathcal{K}}\lambda_k   \phi\ijkC\left(y, \mathcal{S}_j\Co(y)\right) \! \right], \nonumber  \\
%\end{align}
%}
%where $\phi\ijkC(y, s)$ is given by
%\begin{align}
\phi\ijkC(y, s)  &=\sum_{\Tin} \left[\int_{\hik}^{R'} x \rho\ikTo(x) dx + \right. \nonumber \\
& \left. \int_{R'}^\infty x\rho\ikTo(x)\left(1-\frac{1}{1+ s P_k x^{-\alpha\To}}\right) dx \right],\nonumber
\end{align}%
and $R'=\max\left(\RI(y),\;\hik\right)$,}
Notice that $\phi\ijkC(y,s)$ is increase with $y$ and $s$.
%we replace $\hat\alpha\coL=\alpha\Co/\alpha\coL$ and $\hat\alpha\coN=\alpha\Co/\alpha\coN$/
In addition, we can derive the differential of the total \ac{STP} with density $\lambda_j$ as
\begin{align}
 \frac{\partial}{\partial \lambda_j} \mathcal{P}&=\sum_{k\in\mathcal{K}}\int_{\hik}^{\infty} \frac{\partial}{\partial \lambda_j}\left(\varphi_{k}\coL(y) + \varphi_{k}\coN(y) \right)dy\label{eq:A_bound_4}
\end{align}
where components inside the integral are given by
\begin{align}
\frac{\partial}{\partial \lambda_j}  \varphi\ijCo(y)&= \frac{\varphi\ijCo(y)}{\lambda_j}\left(1  -2\pi\lambda_j \phi_{j,j}\Co(y, \mathcal{S}\ijCo(y))   \right)   \label{eq:A_bound_5} \\
\frac{\partial}{\partial \lambda_j}  \varphi_{j'}\Co(y)&= \varphi_{j'}\Co(y)\left(  -2\pi \phi_{j',j}\Co(y, \mathcal{S}_j\Co(y))  \right) \label{eq:A_bound_6}
\end{align}
where $j'$ used to represent $j'\neq j$.
Notice that \eqref{eq:A_bound_5} is differential of the \ac{STP} when main link is $j$-layer. On the other hand, \eqref{eq:A_bound_6} is differential with $\lambda_j$ for the \ac{STP} when main link is not $j$-layer, hence the \ac{STP} always decrease with $\lambda_k$.

%Since the total \ac{STP} and the differential is not closed form, which makes hard to find the optimal densities, we derive range of $\lambda_j$ that makes differential of \ac{STP} negative to derive the upper bound of optimal $\lambda_j$.
From \eqref{eq:A_bound_5} and \eqref{eq:A_bound_6}, we derive the range of $\lambda_j$ that makes the total \ac{STP} decreased with the $\lambda_j$.
Here, $\varphi_{j}\Co(y)$ and $\lambda_j$ is positive for all domain.
Hence, the total \ac{STP} decreases with $\lambda_j$, if following inequality holds.
\begin{align}
 \underset{y, c}{\max} &\left[\frac{1}{2\pi   \phi_{j,j}\Co(y, \mathcal{S}_j\Co(y))    }\right] \leq \lambda_j.\label{eq:A_bound_7} % \implies \frac{\partial}{\partial \lambda_j}\mathcal{P}_i\leq 0 \label{eq:A_bound_7} \\
\end{align}
Furthermore, as $\phi_{j,j}\Co(y,s)$ is increase with $y$ and $s$, we can put the minimum value of $y$ and $s$ which gives
\begin{align}
\phi_{j,j}\coL \left(  \hij, S_j\coL(\hij)\right)=  \epsilon_j\left(\mathcal{S}_j\coL(\hij)\right)  .\label{eq:A_bound_8}
\end{align}
Hence, the upper bound of optimal density is given by \eqref{eq:L_Up_1}.
%\begin{align}
%\lambda_j^b   =\frac{1}{2\pi   \epsilon_j\left(\mathcal{S}_j\coL(\hij)\right)    } .\label{eq:A_bound_9}
%\end{align}

%%%%%%%%%%%%%%%%%%%%%%%%%%%%%%

\end{appendix}

\bibliographystyle{IEEEtran}

\bibliography{./Bib/StringDefinitions,./Bib/IEEEabrv,./Bib/myBibDongsun}

\end{document}